\documentclass[11pt]{article}
\usepackage{amssymb,amsmath,amsfonts}
\usepackage{graphicx}
\usepackage{graphics}
\usepackage{eepic,epsfig}
\usepackage{verbatim}
\usepackage{color}
\usepackage{hyperref}
1.60cm \makeatletter \@addtoreset{equation}{section}

\makeatother

\begin{document}
\title{Induced vacuum bosonic current by magnetic flux in a higher dimensional 
compactified cosmic string spacetime}
\author{E. A. F. Bragan\c{c}a$^1$\thanks{E-mail:eduardoandre1983@gmail.com} ,
H. F. Santana Mota$^2$\thanks{E-mail: hm288@sussex.ac.uk} and E. R. Bezerra de Mello$^3$\thanks
{E-mail: emello@fisica.ufpb.br}\\
\\
\textit{$^{1,3}$Departamento de F\'{\i}sica, Universidade Federal da Para\'{\i}ba}\\
\textit{58.059-970, Caixa Postal 5.008, Jo\~{a}o Pessoa, PB, Brazil}\vspace{%
0.3cm}\\
\textit{$^{2}$Department of Physics and Astronomy, University of Sussex}\\
\textit{Falmer,Brighton BN1 9QH, U.K.}}
\maketitle
%
\begin{abstract}
%
In this paper, we analyse the bosonic current densities induced by a 
magnetic flux running along an idealized cosmic string in a
high-dimensional spacetime, admitting that the coordinate along the string's
axis is compactified. Additionally we admit the presence of an
magnetic flux enclosed by the compactification axis. 
In order to develop this analysis we calculate
the complete set of normalized bosonic wave-functions obeying a
quasiperiodicity condition, with arbitrary phase $\beta$, along the compactified dimension. In this context, 
only azimuthal and axial currents densities take place. As to the azimuthal current, two contributions appear. 
 The first contribution corresponds to the standard azimuthal
current in a cosmic string spacetime without compactification, while the
second contribution is a new one, induced by the
compactification itself. The latter is an even
function of the magnetic flux enclosed by the string axis and is an odd function 
of the magnetic flux along its  core with period equal to quantum flux, $\Phi_0=2\pi/e$.
On the other hand, the 
nonzero axial current density is an even function 
of the magnetic flux along the core of the string and an odd function 
of the magnetic flux enclosed by it. We also find that the
axial current density vanishes for untwisted and twisted bosonic
fields in the absence of the magnetic flux enclosed by the string axis.
Some asymptotic expressions for the current density are provided for
specific  limiting cases of the physical 
parameter of the model.
\end{abstract}
\bigskip


\bigskip
%
\section{Introduction}
\label{Int}
%
Cosmic strings are linear gravitational stable topological defects which may have
been created as a consequence of phase transitions in the early universe and are predicted
in the context of the standard gauge field theory of elementary particle physics
with extra symmetries \cite{VS,hindmarsh,Hyde:2013fia}. Observations of anisotropies
in the Cosmic Microwave Background Radiation (CMB) by COBE, WMAP and more recently by the Planck Satellite have ruled out
cosmic strings as the primary source for primordial density perturbations since they cannot explain
the acoustic peaks in the angular power spectrum. However, a couple of other effects due to
cosmic strings can be investigated through the observations of CMB, such as non-gaussianity
signals and temperature anisotropies caused by the gravitational field of moving strings
(see \cite{Ade:2013xla} for a recent analysis of the possible effects of strings on CMB).
Furthermore, the formation of strings can also have astrophysical and cosmological consequences.
For instance, emission of gravitational waves and high energy cosmic rays by strings
such as neutrinos and gamma-rays, along with observational data, can help to constraint
the product of the Newton's constant, $G$, and the linear mass density of the string
$\mu_0$ \cite{hindmarsh}. All these effects can also be generated by cosmic strings formed
in the context of brane inflation\footnote{For a review of the observational consequences generated
by cosmic strings see \cite{Hindmarsh:2011qj, Copeland:2011dx}.}, which makes the physics of cosmic strings
a wide area of worth investigation.  
In this paper, we will only focus on the effects generated by
the spacetime of a infinitely long straight static cosmic string as explained below. 

The geometry of the spacetime associated with an idealized
cosmic string,  i.e., infinitely long and straight,  is locally flat
but topologically conical, having a planar angle deficit given by $\Delta\phi=8\pi G\mu_0$
on the two-surface orthogonal to the string.
Although this object was first 
introduced in the literature as being created by a Dirac-delta type
distribution of energy and axial stress  along a straight infinity and
line, it can also  be described by classical field theory where
the energy-momentum tensor associated with the Maxwell-Higgs 
system, investigated by Nielsen and Olesen in
\cite{Nielsen197345}, couples to the Einstein's equations. This coupled system was further
investigated by Garfinkle and Linet in \cite{PhysRevD.32.1323} and \cite{Linet1987240},
respectively. These authors have shown that a planar angle deficit, $\Delta\phi $, arises
on the two-surface perpendicular to a string, as well as a magnetic
flux running through its core.

Although the geometry of the spacetime produced by an idealized cosmic
string is locally flat, its conical structure alters the vacuum fluctuations associated with quantum fields.
As a consequence,  the vacuum expectation value (VEV)  of physical observables like
the energy-momentum tensor, $\langle T_{\mu\nu}\rangle$, gets a nonzero value.
The calculation of the VEV of physical observables associated with the scalar and fermionic
fields in the cosmic string spacetime has
been developed in  \cite{PhysRevD.35.536, escidoc:153364, GL, DS, PhysRevD.46.1616}
and \cite{PhysRevD.35.3779, LB, Moreira1995365, BK}, respectively. Furthermore, the presence of
a magnetic flux running through the core of the string gives additional
contributions to the VEVs associated with charged fields
\cite{PhysRevD.36.3742, guim1994, SBM, SBM2, SBM3, Spinelly200477, SBM4} as well as
induces vacuum current densities, $\langle j^\mu\rangle$. 
This phenomenon has been investigated for massless and massive scalar 
fields in \cite{LS} and \cite{SNDV}, respectively. In these papers, 
the authors have shown that induced vacuum
current densities along the azimuthal direction arise if the ratio of the
magnetic flux by the quantum one has a nonzero fractional part. The calculation
of induced fermionic currents in higher-dimensional cosmic string spacetime
in the presence of a magnetic flux has been developed in \cite{ERBM}.
The induced fermionic current by a magnetic flux in $(2+1)$-dimensional 
conical spacetime and in the presence of a circular boundary has
also been analyzed in \cite{PhysRevD.82.085033}. 

The presence of compact dimensions also induces topological quantum
effects on matter fields. It is well known that the presence of compact 
dimensions is an important feature in many high-energy theories of fundamental physics,
like supergravity and superstring theories. An interesting application
of field theoretical models that present compact dimensions can be found in
nanophysics. The long-wavelength description of the electronic states in
graphene can be formulated in terms of the Dirac-like theory in
three-dimensional spacetime, with the Fermi velocity playing the role of the
speed of light (see, e.g., \cite{RevModPhys.81.109}). The corresponding
effective $(2+1)-$ dimensional field theory, in addition to Dirac fermions, involves scalar
and gauge fields originated from the elastic properties and describing disorder phenomena, like
the distortion of graphene lattice and structural defects \cite{Jackiw:2007rr,Oliveira:2010hq}.
  For instance, a single-walled carbon nanotube
is generated by rolling up a graphene sheet to form a cylinder. In this case, the
background spacetime for the corresponding Dirac-like theory has topology $
R^{2}\times S^{1}$. The combined effects of the non-trivial topology of
the cosmic string spacetime, the compactified dimension along the axis of the string and
the presence of a magnetic flux running through its core and enclosed by
the compactified axis, on the VEVs of the energy-momentum tensor, $\langle T_{\mu\nu}\rangle$,
and current densities, $\langle j_\mu\rangle$, associated with charged quantum
fermionic fields in a four-dimensional cosmic string spacetime have recently been investigated in \cite{SERA} and
\cite{BMSAA}, respectively. In Kaluza-Klein type models, the currents along compact dimensions
generate magnetic filed in the uncompactified subspace. The magnetic
fields induced by the vacuum currents provide an additional channel for interaction
of cosmic string with the enviroment. 
 Here, in the present paper, we shall continue along
the same line of investigation. We shall calculate the induced bosonic current
in a higher-dimensional cosmic string spacetime, under
the same conditions as considered in these two previous publications. 

This paper is organized as follows. The section \ref{Wightman} is devoted
to the evaluation of the positive frequency Wightman function for a massive
charged scalar quantum field in a higher-dimensional cosmic string spacetime. We also consider that the $z$-axis
along the string is compactified to a circle, by imposing a quasiperiodic boundary condition
on the bosonic field with arbitrary phase. Moreover we assume the presence of  
magnetic fluxes running through the string's core and enclosed by its axis.
In section \ref{current}, by using the Wightman function, 
we evaluate the renormalized vacuum current
density induced by the magnetic fluxes and the compactification. As we
shall see, the renormalized charge density and the radial current vanish. 
For the azimuthal current density, the compactification induces it to 
decompose into two parts: one of them coincides with the corresponding
expression in the geometry of a cosmic string without compactification
and the other is the contribution due to the compactification itself.
Moreover, as a consequence of the compactification, a
non-vanishing axial current also arises and is a purely topological
one.
The most relevant conclusions of the paper are summarized in section \ref{conc}.
We have also dedicated an Appendix to provide some important expressions 
used in the development of our calculation for the induced current densities.
Throughout the paper we use natural units $G=\hbar =c=1$.
%
\section{Wightman function}
\label{Wightman}
%

In this paper we consider a $(D+1)-$dimensional idealized cosmic string spacetime
with $D\geq 3$. By using
the generalized cylindrical coordinates $(x^1,x^2,...,x^D)=(r,\phi,z,x^4,...,x^D)$ with 
the string on the $(D-2)-$dimensional hypersurface $r=0$, the corresponding geometry 
is described by the line element
\begin{equation}
ds^{2}=g_{\mu\nu}dx^{\mu}dx^{\nu}=dt^{2}-dr^2-r^2d\phi^2-dz^2- \sum_{i=4}^{D}(dx^{i})^2 \ .
\label{eq01}
\end{equation}

The coordinates take values in the following intervals: $r\geq 0$, $0\leq\phi\leq 2\pi/q$ and 
$-\infty< (t, \ x^i) < +\infty$ for $i=4,...,D$.  The parameter $q\geq 1$ codifies the presence of the
cosmic string. Moreover, we assume that the direction
along the $z$-axis is compactified to a circle with the length $L$, so $0\leqslant z\leqslant L$.
The standard cosmic string space-time is characterized by $D=3$, with $z\in(-\infty, \ \infty)$.
In this case $q^{-1}=1-4\mu_0$, being $\mu_0$ the linear mass density of the string.\footnote{
It is interesting to note that the effective metric produced in superfluid $
^{3}\mathrm{He-A}$ by a radial disgyration is described by $D=3$ line
element (\ref{eq01}) with a negative planar angle deficit \cite{Volovik}.} 

In this paper we are interested in calculating the induced vacuum current density, $\langle j_{\mu}\rangle$, associated with
a charged scalar quantum field, $\varphi (x)$, in the presence of
magnetic fluxes running along the core of the string and enclosed by it,
considering that the $z$-axis is compactified to a circle. In order to
do that we shall calculate the corresponding complete set of normalized 
bosonic wavefunction. 

The equation which governs the quantum dynamics of a charged
bosonic field with mass $m$, in a curved background and in the presence of an 
electromagnetic potential vector, $A_\mu$, reads
\begin{equation}
\left({\cal D}^2+m^{2}+\xi R\right) \varphi (x)=0 \ ,
\label{eq02}
\end{equation}
where the differential operator above is defined by
\begin{eqnarray}
{\cal D}^2=\frac{1}{\sqrt{|g|}}D_\mu\left(\sqrt{|g|}\,g^{\mu\nu}D_\nu\right) \ ,
\end{eqnarray}
being $D_{\mu}=\partial_{\mu}+ieA_{\mu}$ and $g={\rm det}(g_{\mu\nu})$. In addition,
we have considered the presence of a non-minimal coupling, $\xi$, between the field and the
geometry represented by the Ricci scalar, $R$. However, for a thin and infinitely straight cosmic string, $R=0$ for $r\neq 0$. 

In the analysis that we want to develop, it will be assumed that the
direction along the $z$-axis is compactified to a circle with length $L$: $
0\leqslant z\leqslant L$. The compactification is achieved by imposing the
quasiperiodicity condition on the matter field,
\begin{equation}
\varphi (t,r,\phi,z+L,x^4,...,x^D)=e^{2\pi i\beta}\varphi(t,r,\phi,z,x^4,...,x^D) \ ,  
\label{eq03}
\end{equation}
with a constant phase $\beta $, $0\leqslant \beta \leqslant 1$. The special
cases $\beta =0$ and $\beta =1/2$ correspond to the untwisted and twisted
fields, respectively, along the $z$-direction. In addition, we shall consider the
existence of the following constant vector potential
\begin{equation}
A_{\mu}=(0,0,A_{\phi},A_{z})\ , 
\label{eq05}
\end{equation}
with $A_{\phi}=-q\Phi_\phi/(2\pi)$ and $A_{z}=-\Phi_z/L$, being $\Phi_\phi$ and $\Phi_z$
the corresponding magnetic fluxes. In quantum field  theory the  condition
\eqref{eq03} changes the spectrum of the
vacuum fluctuations compared with the case of uncompactified dimension and,
as a consequence, the induced vacuum current density changes. 

In the spacetime defined by \eqref{eq01} and in the presence of the vector 
potential given above, the equation \eqref{eq02} becomes
\begin{equation}
\left[\partial_t^2-\partial_r^2-\frac{1}{r}\partial_r-\frac{1}{r^2}(\partial_{\phi}+
ieA_{\phi})^2-(\partial_{z}+ieA_{z})^2-\sum_{i=4}^{D}\partial_{i}^{2}+m^2\right]\varphi(x)=0 \ . 
\label{eq06}
\end{equation}

The positive energy solution of  this equation can be obtained 
by considering the general expression,
\begin{equation}
\varphi(x)=CR(r)e^{-i\omega t+iqn\phi+ik_z z+i{\vec{k}\cdot{\vec{r}}_{\parallel}}}  \ ,
\label{eq07}
\end{equation}
where ${\vec{r}}_{\parallel}$ represents the coordinates of the extra dimensions,
$\vec{k}$ the momentum along these directions and $C$ is a normalization constant. 
Substituting \eqref{eq07} into \eqref{eq06} 
we find that the radial function $R(r)$ must obey the differential equation
\begin{equation}
\left(\frac{d^2}{dr^2}+\frac{1}{r}\frac{d}{dr}+\lambda^2-\frac{\nu^2}{r^2}\right)R(r)=0 \ , 
\label{eq08}
\end{equation}
with
\begin{eqnarray}
\lambda&=&\sqrt{\omega^2-{\vec{k}}^{2}-{\tilde{k}}_z^2-m^2} \ , \nonumber\\  
\nu&=&qn+eA_{\phi} \ , \nonumber\\  
{\tilde{k}}_z&=&k_z+eA_z \ .
\label{eq09}
\end{eqnarray}

In the present analysis we shall assume that the wave functions obey the Dirichlet boundary condition on the string's core.
The regular solution at origin is $R(r)=J_{\nu}(\lambda r)$, where $J_\nu(z)$ 
represents the Bessel function of order  
\begin{equation}
\nu=\nu_n=q|n+\alpha| \ \ {\rm with} \ \alpha=\frac{eA_\phi}{q}=-\frac{\Phi_\phi}{\Phi_0} \ , 
\label{alpha}
\end{equation}
$\Phi_0=2\pi/e$ being the quantum flux. Then, the general solution takes the form
\begin{equation}
\varphi_{\sigma}(x)=CJ_{q|n+\alpha|}(\lambda r)
e^{-i\omega t+iqn\phi+ik_z z+i{\vec{k}\cdot{\vec{r}}_{\parallel}}} \ . 
\label{eq10}
\end{equation}

The quasiperiodicity condition \eqref{eq03} provides a discretization 
of the quantum number $k_z$ as shown below:
\begin{equation}
k_z=k_l=\frac{2\pi}{L}(l+\beta) \ \ {\rm with} \  l=0\ ,\pm 1,\pm 2,... \ .
\label{eq11}
\end{equation}

Under this circumstance the energy takes the form
\begin{equation}
\omega=\omega_l=\sqrt{m^2+\lambda^2+{\tilde{k}}^2_l+{\vec{k}}^2} \ , 
\label{eq12}
\end{equation}
where
\begin{eqnarray}
{\tilde{k}}_z&=&{\tilde{k}}_l=\frac{2\pi}{L}(l+\tilde{\beta}) \ , \nonumber\\
\tilde{\beta}&=&\beta+\frac{eA_zL}{2\pi}=\beta-\frac{\Phi_z}{\Phi_0} \ .
\label{eq13}
\end{eqnarray}

The constant $C$ can be obtained by the normalization condition
\begin{equation}
i\int d^Dx\sqrt{|g|}\left[\varphi_{\sigma'}^{*}(x)\partial_t
\varphi_{\sigma}(x)-\varphi_{\sigma}(x)\partial_t\varphi_{\sigma'}^{*}(x)
\right]=\delta_{\sigma,\sigma'} \ ,
\label{eq14}
\end{equation}
where the delta symbol on the right-hand side is understood as Dirac delta 
function for the continuous quantum number, $\lambda$ and ${\vec{k}}$, and
Kronecker delta for the discrete ones, $n$ and $k_l$. From \eqref{eq14} one finds 
\begin{equation}
|C|^2=\frac{q\lambda}{2(2\pi)^{D-2}\omega_l L} \ .
\label{eq15}
\end{equation}

So, the renormalized bosonic wave-function reads,
\begin{equation}
\varphi_{\sigma}(x)=\left[\frac{q\lambda}{2(2\pi)^{D-2}\omega_l L}\right]^{\frac{1}{2}} 
J_{q|n+\alpha|}(\lambda r)e^{-i\omega t+iqn\phi+ik_l z+i{\vec{k}\cdot{\vec{r}}_{\parallel}}} \ .
\label{eq16}
\end{equation}

The properties of the vacuum state are described by the corresponding positive 
frequency Wightman function, $W(x,x')=\left\langle 0|\varphi(x)\varphi^{*}(x')|0 \right\rangle$, 
where $|0 \rangle$ stands for the vacuum state. Having this function we can evaluate
the induced bosonic current. For the evaluation of the Wightman 
function, we use the mode sum formula
\begin{equation}
W(x,x')=\sum_{\sigma}\varphi_{\sigma}(x)\varphi_{\sigma}^{*}(x') \ ,
\label{eq17}
\end{equation}
where we are using the compact notation defined as
\begin{equation}
\sum_{\sigma }=\sum_{n=-\infty}^{+\infty} \ \int d{\vec{k}} \ \int_0^\infty
\ d\lambda \ \sum_{l=-\infty }^{+\infty} \ .  \label{Sumsig}
\end{equation}
The set $\{\varphi_{\sigma}(x), \ \varphi_{\sigma}^{*}(x')\}$ represents a complete set 
of normalized mode functions satisfying the periodicity condition \eqref{eq03}.
In our case, the  mode functions in Eq. \eqref{eq16} are specified by the set of quantum numbers 
$\sigma=(n,\lambda,k_l, {\vec{k}})$, with the values in the ranges $n=0,\pm1,\pm2, \ \cdots$, 
$-\infty<k^j<+\infty$ with $j=4, \cdots \ D$, $0<\lambda<\infty$ and $k_l=2\pi(l+\beta)/L$
with $l=0,\pm1,\pm2, \ \cdots$. 

Substituting \eqref{eq16} into the sum \eqref{eq17} we obtain
\begin{eqnarray}
W(x,x')&=&\frac{q}{2L(2\pi)^{D-2}}\sum_\sigma e^{iqn\Delta\phi} 
 e^{i{\vec{k}\cdot{\Delta\vec{r}}_{\parallel}}} \ 
\lambda J_{q|n+\alpha|}(\lambda r) J_{q|n+\alpha|}(\lambda r')\nonumber\\
&\times&\frac{e^{-i\omega_l\Delta t+ik_l\Delta z}}{\omega_l} \ ,
\label{eq18}
\end{eqnarray}
where $\Delta\phi=\phi-\phi'$, $\Delta{\vec{r}}_{\parallel}={\vec{r}}_{\parallel}-
{\vec{r}}_{\parallel}'$, $\Delta t=t-t'$ and $\Delta z=z-z'$. 

Having the positive frequency Wightman function above, we are in position
to calculate the induced vacuum  bosonic current density, $\langle j_\mu\rangle$.
This calculation will be developed in the next section.
%
\section{Bosonic current}
\label{current}
%
The bosonic current density operator is given by,
\begin{eqnarray}
j_{\mu }(x)&=&ie\left[\varphi ^{*}(x)D_{\mu }\varphi (x)-
(D_{\mu }\varphi)^{*}\varphi(x)\right] \nonumber\\
&=&ie\left[\varphi^{*}(x)\partial_{\mu }\varphi (x)-\varphi(x)
(\partial_{\mu }\varphi(x))^{*}\right]-2e^2A_\mu(x)|\varphi(x)|^2 \   .
\label{eq20}
\end{eqnarray}
Its vacuum expectation value (VEV) can be evaluated in terms of the positive frequency Wightman 
function as shown below:
\begin{equation}
\left\langle j_{\mu}(x) \right\rangle=ie\lim_{x'\rightarrow x}
\left\{(\partial_{\mu}-\partial_{\mu}')W(x,x')+2ieA_\mu W(x,x')\right\} \ .
\label{eq21}
\end{equation}

This VEV is a periodic function of the magnetic fluxes $\Phi_\phi$ and
$\Phi_z$ with period equal to the quantum flux. This can be observed if we write 
the parameter $\alpha$ in \eqref{alpha} in the form
\begin{equation}
\alpha = n_{0}+\alpha_{0}  \ {\rm with} \ |\alpha_{0}|<\frac{1}{2} \ ,
\label{alphazero}
\end{equation}
where $n_{0}$ is an integer number. In this case the VEV of the current density will
depend on $\alpha_{0}$ only.
%
\subsection{Charge density and radial current}
%
Let us start the calculation with the charge density. Because $A_0=0$, we have,
\begin{equation}
\rho(x)=\left\langle j^{0}(x) \right\rangle= ie\lim_{x' \rightarrow x}
(\partial_{t}-\partial_{t'})W(x,x') \ .
\label{chargedensity}
\end{equation}

Substituting \eqref{eq18} into the above expression, taking the time derivative 
and finally the coincidence limit $x' \rightarrow x$, the formal expression for the
charge density is:
\begin{equation}
\rho(x)=\frac{q e}{(2\pi)^{D-2}L}\sum_\sigma \ \lambda \ J^2_{q|n+\alpha|}(\lambda r) \ .
\label{density}
\end{equation}

Because the above expression is divergent, in order to obtain a finite 
and well defined result, we have to regularize 
it by introducing a cutoff function  $e^{-\eta(\lambda^{2}+{\vec{k}}^{2}+k_l^2)}$,
with the cutoff parameter $\eta>0$. At the end of the calculation we shall take 
the limit $\eta\rightarrow0$.

So, using the cutoff function, the integral over the variable $\lambda$
can be evaluated with the help of \cite{gradshteyn2000table}, and the regularized 
contribution gives
\begin{equation}
\int_{0}^{\infty}d\lambda \ \lambda e^{-\eta \lambda^{2}}J^{2}_{q|n+\alpha|}
(\lambda r)=\frac{1}{2\eta}e^{-\frac{r^{2}}{2\eta}}I_{q|n+\alpha|}
\left({r^2}/(2\eta)\right) \ ,
\label{intJ}
\end{equation}
with $I_{\nu}(z)$ being the modified Bessel function. As to the integral over 
${\vec{k}}$ we have,
\begin{equation}
\int \ d{\vec{k}} \ e^{-\eta {\vec{k}}^2}=\left(\frac{\pi}{\eta}\right)^{\frac{(D-3)}{2}} \ .
\label{intK}
\end{equation}

Thus, the regularized charge density reads
\begin{equation}
\rho_{{\rm reg}}(x,\eta)=\frac{qe}{(4\pi)^{\frac{(D-1)}{2}}L}\frac{e^{-\frac{r^{2}}{2\eta}}}{\eta^{\frac{(D-1)}{2}}}
\sum_{l=-\infty}^\infty e^{-\eta  k_l^2} {\cal{I}}(q,\alpha_0,r^2/(2\eta))\ ,
\label{densityreg}
\end{equation}
where
\begin{equation}
{\cal{I}}(q,\alpha_0,w)=\sum_{n=-\infty}^\infty I_{q|n+\alpha|}(w)=
\sum_{n=-\infty}^\infty I_{q|n+\alpha_0|}(w) \ .
\label{sum1}
\end{equation}

In Appendix \ref{summ1}, it is shown
that the above summation on the quantum number $n$
is given by \eqref{eq:10} and \eqref{eq:09}. Substituting
this result into \eqref{densityreg} we obtain
\begin{eqnarray}
\rho_{{\rm reg}}(x,\eta)&=&\frac{qe}{(2\pi)^{\frac{(D-1)}{2}}L}\frac{w^{\frac{(D-1)}{2}}e^{-w}}{r^{(D-1)}}
\sum_{l=-\infty}^\infty e^{-\frac{r^2k_l^2}{2w}}\left[
 \frac{e^{w}}{q}-\frac{1}{\pi}\right.
\int_{0}^{\infty}dy\frac{e^{-w\cosh(y)}f(q,\alpha_{0},y)}{\cosh(qy)-\cos(q\pi)}\nonumber\\
&+&\left.\frac{2}{q}\sideset{}{'}\sum_{k=1}^{p}\cos(2k\pi\alpha_{0})e^{w\cos(2k\pi /q)}\right] \ ,
\label{densityreg1}
\end{eqnarray}
with $w=r^2/(2\eta)$.  In the above expression $p=[q/2]$, 
where $[q/2]$ represents the integer part of $q/2$, and the prime on the sign of
the summation means that in the case $q=2p$ the term $k=q/2$ should be
taken with the coefficient $1/2$. 

 The first term in the square bracket of (\ref{densityreg1}) 
corresponds to the charge density for $\alpha _{0}=0$ and $q=1$. 
The renormalized value for the charge density is given by subtracting from
\eqref{densityreg1} the contribution corresponding to the Minkowski spacetime 
in the absence of magnetic flux. We can do this in manifest form by discarding the first term
inside the bracket.
The other contributions contain $e^{-2w
\cosh ^{2}(y/2)}$ and $e^{-2w\sin^{2}(\pi k/q) }$, inside 
the integral and summation respectively; hence
in the limit $\eta \rightarrow 0$ ($w \rightarrow\infty$) these terms vanish for $r>0$. Thereby, we
conclude that the renormalized value for the charge density is zero, i.e,
there is no induced charge density.

In the Apendix \ref{extra-dimension} we explicitly proved that  the renormalized induced vacuum current densities
along the extra dimensions vanish, i.e., $\langle j_i(x)\rangle_{ren}=0$, for $i=4, \ ... \ D$. 
This result is in agreement with the invariance of the system under a boost along the $x^i$ direction.

Now let us analyze the radial current density.
Because $A_r=0$, the VEV of the $r$-component of the current is simply expressed as
\begin{eqnarray}
\left\langle j_{r}(x) \right\rangle= ie\lim_{x'\rightarrow x}
(\partial_{r}-\partial_{r'})W(x,x') \ .
\end{eqnarray}
Taking the radial derivatives with respect to
$r$ and $r'$ in the Wightman function, subtracting both terms 
and taking the coincidence limit, 
there appears a cancellation between those terms. Thereby, we also conclude
that there is no induced radial current density:
\begin{equation}
\left\langle j_{r}(x) \right\rangle=0.
\end{equation}

%
\subsection{Azimuthal current}
%
The VEV of the azimuthal current density is given by
\begin{equation}
\left\langle j_{\phi}(x) \right\rangle = ie \lim_{x '\rightarrow x}
\left\{(\partial_{\phi}-\partial_{\phi '})W(x,x')+2ieA_{\phi}W(x,x')\right\} \ .
\label{jphi}
\end{equation}

Substituting \eqref{eq18} into the above equation 
we get the formal expression for the azimuthal bosonic current density below:
\begin{eqnarray}
\left\langle j_{\phi}(x) \right\rangle&=&-\frac{qe}{L(2\pi)^{D-2}}\sum_{n=-\infty}^\infty
q(n+\alpha)\int \ d{\vec{k}}\int_0^\infty \ d \lambda \ \lambda \ J^2_{q|n+\alpha|}(\lambda r)\nonumber\\ 
&\times&\sum_{l=-\infty}^\infty\frac{1}{\sqrt{m^2+\lambda^2+{\tilde{k}}^2_l+{\vec{k}}^2}} \ .
\label{jphi1}
\end{eqnarray}

In order to develop the summation over the quantum number $l$ we shall apply
the Abel-Plana summation formula in the form \cite{PhysRevD.82.065011}, which is given by
\begin{eqnarray}
&&\sum_{l=-\infty }^{\infty }g(l+\tilde{\beta} )f(|l+\tilde{\beta} |)=\int_{0}^{\infty }du\,
\left[ g(u)+g(-u)\right] f(u)  \notag \\
&&\qquad +i\int_{0}^{\infty }du\left[ f(iu)-f(-iu)\right] \sum_{\lambda =\pm
1}\frac{g(i\lambda u)}{e^{2\pi (u+i\lambda \tilde{\beta} )}-1} \ .
\label{sumform}
\end{eqnarray}
Taking $g(u)=1$ and 
\begin{equation}
f(u)=\frac{1}{\sqrt{(2\pi u/L)^{2}+\lambda^2+m^{2}+
{\vec{k}}^{2}}} \ .    \label{fg}
\end{equation}
By using this formula, it is possible to decompose the expression
to $\langle j_{\phi}\rangle$, Eq. \eqref{jphi1}, as the sum of the
two contributions as shown below:
\begin{equation}
\langle j_{\phi}\rangle =\langle j_{\phi }\rangle _{{\rm cs}}+\langle j_{\phi
}\rangle _{{\rm c}} \ ,
\label{total}
\end{equation}
where the term, $\langle j^{\phi }\rangle _{{\rm cs}}$, is the contribution of
the first integral on the right-hand side of (\ref{sumform}) and corresponds
to the azimuthal current density in the geometry of the 
higher dimensional cosmic string spacetime without
compactification. The term, $\langle j^{\phi }\rangle _{{\rm c}}$, is induced by the
compactification of the string along its axis and is provided by the second integral on the right-hand 
side of (\ref{sumform}). As we shall see the latter
vanishes in the limit $L\rightarrow \infty $.

As we have already mentioned in the beginning of this paper, 
the calculation of the induced azi\-muthal bosonic current density by a magnetic
flux in the geometry of an idealized cosmic string has been developed 
in \cite{LS} and \cite{SNDV} for massless and massive quantum fields, respectively.
In \cite{SNDV} the calculation of the azimuthal vacuum current density was developed
for a higher dimensional cosmic string spacetime considering the case where
the parameter $1\leq q<2$. However, to our knowledge, a closed
expression for the induced azimuthal current considering general values of
$q$ is missed. In this sense, here we generalize the results in \cite{SNDV} considering 
general values of the parameter, $q$, as well as consider the case where there exists a magnetic flux running through the
core of a string in which the axis is compactified to a circle.
Thus, in the present paper, we want 
to investigate the induced bosonic current as 
general as possible, combining all the above effects.

From the first integral on the right hand side of \eqref{sumform}, Eq.  \eqref{jphi1} gives
\begin{equation}
\left\langle j_{\phi}(x) \right\rangle _{{\rm cs}}= -\frac{2eq}{(2\pi)^{D-1}} 
\sum_{n=-\infty}^{\infty}q(n+\alpha)\int d{\vec{k}}\int_{0}^{\infty}d\lambda \ 
\lambda J_{q|n+\alpha|}^{2}(\lambda r)\int_{0}^{\infty}\frac{dy}{\sqrt{y^{2}+
\lambda^{2}+{\vec{k}}^{2}+m^{2}}} \ ,
\label{jphi_cs02}
\end{equation}
where we have introduced a new variable $y=2\pi u/L$.

In order to provide a more workable expression, we use the identity \eqref{ident}.
This allows us to integrate over $\lambda$ by using 
\cite{gradshteyn2000table}. Also the integral over the momentum on the
extra dimensions is easily evaluated. Finally,  writing $\alpha$ in the 
form given in \eqref{alphazero} we obtain
\begin{equation}
\left\langle j_{\phi}(x) \right\rangle _{{\rm cs}}=-\frac{eq}{(2\pi)^{\frac{(D+1)}{2}}r^{D-1}}\int_{0}^{\infty}
dw \ w^{\frac{(D-3)}{2}}e^{-w-\frac{m^{2}r^{2}}{2w}}\sum_{n=-\infty}^{\infty}q(n+\alpha_{0})
I_{q|n+\alpha_{0}|}(w) \ ,
\label{jphi_cs03}
\end{equation}
where we have defined $w=r^2/2s^2$.

\begin{figure}[!htb]
\begin{center}
\includegraphics[width=0.4\textwidth]{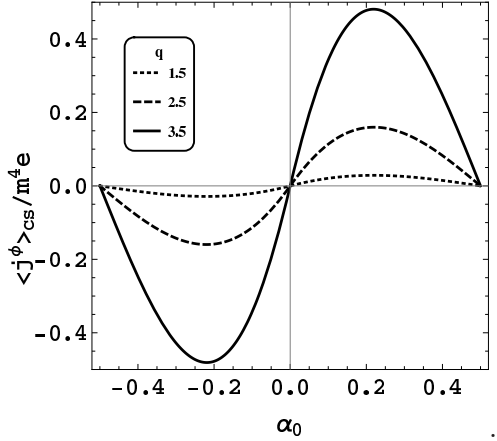}.
\caption{The azimuthal current density without compactification
for $D=3$ is plotted, in units of ``$m^4e$'',
in terms of $\alpha_0$ for $mr=0.5$ and $q=1.5, 2.5$ and $3.5$.}
\label{fig1}
\end{center}
\end{figure}

In Appendix \ref{summ2} it is shown that the above summation
in the quantum number $n$ is given by \eqref{s4} and \eqref{s3}.
Subtituting this result into \eqref{jphi_cs03}, we obtain
\begin{eqnarray}
\left\langle j^{\phi}(x) \right\rangle_{{\rm cs}}&=&\frac{4em^{D+1}}{(2\pi)^{\frac{(D+1)}{2}}}
\left[\sideset{}{'}\sum_{k=1}^{p}\sin(2k\pi/q)\sin(2k\pi\alpha_{0})
 F_{\frac{(D+1)}{2}}\left[2mr\sin(k\pi/q)\right]\right.\nonumber\\
&+&\left.\frac{q}{\pi}
\int_{0}^{\infty}dy \ \frac{g(q,\alpha_{0},2y)\sinh (2y)}{\cosh(2qy)-\cos(q\pi)}F_{\frac{(D+1)}{2}}\left[2mr\cosh(y)\right]
\right],
\label{jphi_cs04}
\end{eqnarray}
where we use the notation
\begin{equation}
F_{\nu}(x)=\frac{K_{\nu}(x)}{x^{\nu}} \ ,
\end{equation}
being $K_\nu(x)$ the modified Bessel function.  We can see that 
$\left\langle j^{\phi}(x) \right\rangle_{{\rm cs}}$ vanishes for the case
$\alpha_0=0$. For $1\leqslant q <2$ the first term on the right hand side
of \eqref{jphi_cs04} is absent and the result coincides
with the one found for the azimuthal induced bosonic current in \cite{SNDV}.
From Eq. \eqref{jphi_cs04}, we can see that $\left\langle j^{\phi}(x) \right\rangle_{{\rm cs}}$
is an odd function of $\alpha_{0}$, with period equal to the quantum flux $\Phi_0$. In Fig.\ref{fig1} we plot the behavior of the
azimuthal current density as a function
of $\alpha_{0}$ for the case where $D=3$, considering $mr=0.5$ and different values of $q$.
One can see that the effect of the cosmic string parameter, $q$, is to
amplify the oscillatory nature of the azimuthal current
with respect to the parameter $\alpha_{0}$.
%
%

For a massless field, from Eq. \eqref{jphi_cs04}, we obtain the following expression
\begin{eqnarray}
\left\langle j^{\phi}(x) \right\rangle_{{\rm cs}}&=&\frac{4e\Gamma\left(\frac{D+1}{2}\right)}{\pi^{\frac{(D+1)}{2}}(2r)^{D+1}}
\left[\sideset{}{'}\sum_{k=1}^{p}\frac{\cos(k\pi/q)
\sin(2k\pi\alpha_{0})}{\sin^{D}(k\pi/q)}\right. \nonumber \\
&+&\left.\frac q\pi\int_{0}^{\infty}dy\frac{\sinh (y)}{\cosh(2qy)-\cos(q\pi)}
\frac{g(q,\alpha_{0},2y)}{\cosh^{D}(y)}\right] \ .
\label{jphi_massless}
\end{eqnarray}
Taking for the above expression $D=3$, we have
\begin{eqnarray}
\left\langle j^{\phi}(x) \right\rangle_{{\rm cs}}&=&\frac{e}{4\pi^{2}r^4}
\left[\sideset{}{'}\sum_{k=1}^{p}\frac{\cot(k\pi/q)\sin(2k\pi\alpha_{0})}{\sin^{2}(k\pi/q)}\right.\nonumber \\
&+&\left.\frac q\pi\int_{0}^{\infty}dy\frac{\tanh (y)}
{\cosh(2qy)-\cos(q\pi)}\frac{g(q,\alpha_{0},2y)}{\cosh^{2}(y)}\right] \ .
\label{jphi_csmasslessd3}
\end{eqnarray}

Considering $D=3$ and $q>2$, the behavior of the $\left\langle j^{\phi}(x) \right\rangle_{{\rm cs}}$
at large distance from string, $mr\gg1$, is dominated by the first term of \eqref{jphi_cs04}
with $k=1$. It reads,
\begin{eqnarray}
\left\langle j^{\phi}(x) \right\rangle_{{\rm cs}}&\approx& \frac{em}{(2\pi r)^2}
\left(\frac m{\pi r \sin(\pi/q)}\right)^{\frac{1}{2}}
\sin(2\pi\alpha_{0})\cot(\pi/q)e^{-2mr\sin(\pi/q)} \ .
\label{jphi_csmrlarged3}
\end{eqnarray}

Now let us develop the calculation of the contribution to the
azimuthal current induced by the compactification. So, we substitute
the second term of \eqref{sumform} into \eqref{jphi1}. Doing this we obtain:
\begin{eqnarray}
\left\langle j_{\phi}(x)\right\rangle_{{\rm c}}&=&-\frac{2eq}{(2\pi)^{D-1}}
\sum_{n=-\infty}^{\infty}q(n+\alpha)\int d{\vec{k}}\int_{0}^{\infty}d\lambda \ 
\lambda J_{q|n+\alpha|}^{2}(\lambda r) \nonumber \\
&&\times\int_{\sqrt{\lambda^{2}+{\vec{k}}^{2}+m^{2}}}^{\infty}\frac{dy}{\sqrt{y^{2}-
\lambda^{2}-{\vec{k}}^{2}-m^{2}}}\sum_{\lambda=\pm1}
\frac{1}{e^{Ly-2\pi i\lambda\tilde{\beta}}-1} \ .
\label{jphi_comp01}
\end{eqnarray}

To continue our analysis, we shall use the series expansion $(e^{u}-1)^{-1}=
\sum_{l=1}^{\infty}e^{-lu}$ in the above expression, and with the help of
\cite{gradshteyn2000table} we get,
\begin{eqnarray}
\left\langle j_{\phi}(x) \right\rangle_{{\rm c}}&=&-\frac{4eq}{(2\pi)^{D-1}}
\sum_{l=1}^{\infty}\cos(2\pi l\tilde{\beta})\sum_{n=-\infty}^{\infty}q(n+\alpha)
\int d{\vec{k}}\int_{0}^{\infty}d\lambda \ \lambda J_{q|n+\alpha|}^{2}
(\lambda r) \times \nonumber \\
&&\times K_{0}\left(lL\sqrt{\lambda^{2}
+{\vec{k}}^{2}+m^2}\right) \ .
\label{jphi_comp02}
\end{eqnarray}

At this point, we shall use the integral representation below for the 
Macdonald function \cite{gradshteyn2000table}
\begin{equation}
K_{\nu}(x)=\frac{1}{2}\left(\frac{x}{2}\right)^{\nu}\int_{0}^{\infty}
dt\frac{e^{-t-\frac{x^{2}}{4t}}}{t^{\nu +1}}.
\label{Macdonald_repres}
\end{equation}

By using this representation, it is
possible to integrate over the variable $\lambda$ and over the
momentum along the extra dimensions ${\vec{k}}$. So, we obtain
\begin{eqnarray}
\left\langle j_{\phi}(x) \right\rangle_{{\rm c}}&=&-\frac{eq}{2^{\frac{(D-3)}{2}}\pi^{\frac{(D+1)}{2}}r^{D-1}}
\sum_{l=1}^{\infty}\cos(2\pi l\tilde{\beta})
\int_{0}^{\infty}dw \ w^{\frac{(D-3)}{2}}
e^{-w\left[1+\frac{l^{2}L^{2}}{2r^{2}}\right]-\frac{r^{2}m^{2}}{2w}}\nonumber \\
&&\times\sum_{n=-\infty}^{\infty}q(n+\alpha_{0})
I_{q|n+\alpha_{0}|}(w) \ ,
\label{jphi_comp03}
\end{eqnarray}
where we have used $\alpha$ in the form \eqref{alphazero} and 
introduced the new variable, $w=\frac{2r^{2}t}{l^{2}L^{2}}$.

The sum over the quantum number
$n$ in \eqref{jphi_comp03} can be developed by using the compact result \eqref{s4}. 
This allows us to perform the integrals over $x$. Our final expression is:
\begin{eqnarray}
\left\langle j^{\phi}(x) \right\rangle_{{\rm c}}&=&\frac{8em^{D+1}}{(2\pi)^{\frac{(D+1)}{2}}}
\sum_{l=1}^{\infty}
\cos(2\pi l\tilde{\beta})\left\{\sideset{}{'}\sum_{k=1}^{p}\sin(2k\pi/q)
\sin(2k\pi\alpha_{0})F_{\frac{(D+1)}{2}}\left[mL\sqrt{l^{2}+\rho_{k}^{2}}\right]\right.\nonumber \\
&&\left.+ \ \frac{q}{\pi}\int_{0}^{\infty}dy\frac{g(q,\alpha_{0},2y) \sinh{(2y)}
 }{\cosh(2qy)-\cos(q\pi)}F_{\frac{(D+1)}{2}}\left[mL\sqrt{l^{2}+\eta^{2}(y)}\right] \right\} \ ,
\label{jphi_comp04}
\end{eqnarray}
where we have defined
\begin{eqnarray}
\rho_{k}=\frac{2r\sin(k\pi/q)}{L} \ , \ \eta(y)=\frac{2r\cosh (y)}{L} \ .
\label{rhoandeta}
\end{eqnarray}

\begin{figure}[!htb]
\begin{center}
\includegraphics[width=0.4\textwidth]{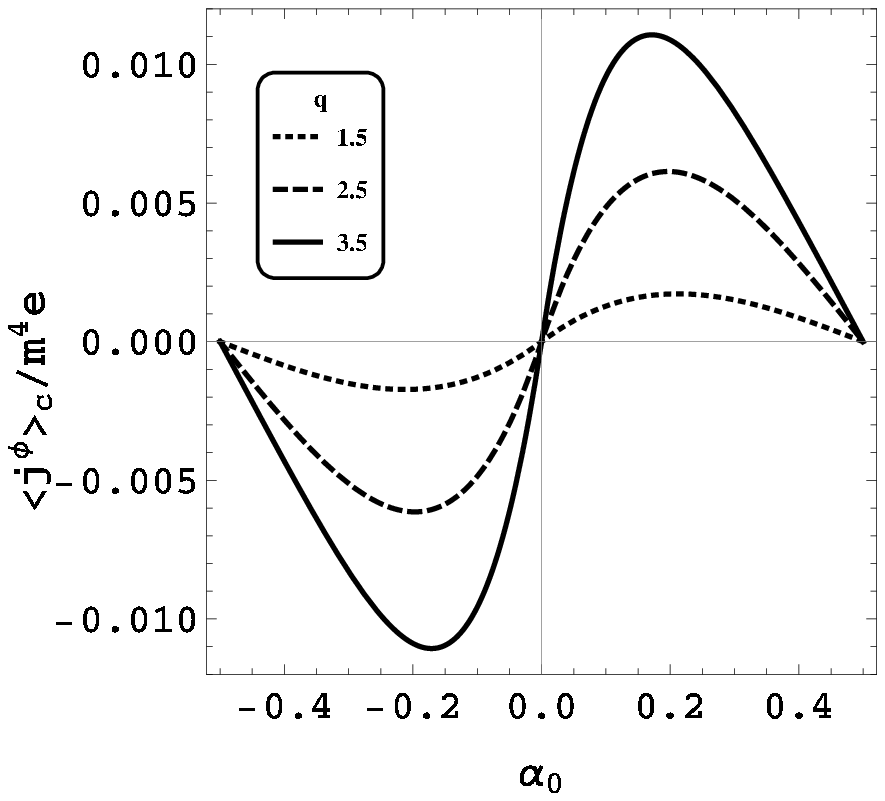}
%
\includegraphics[width=0.4\textwidth]{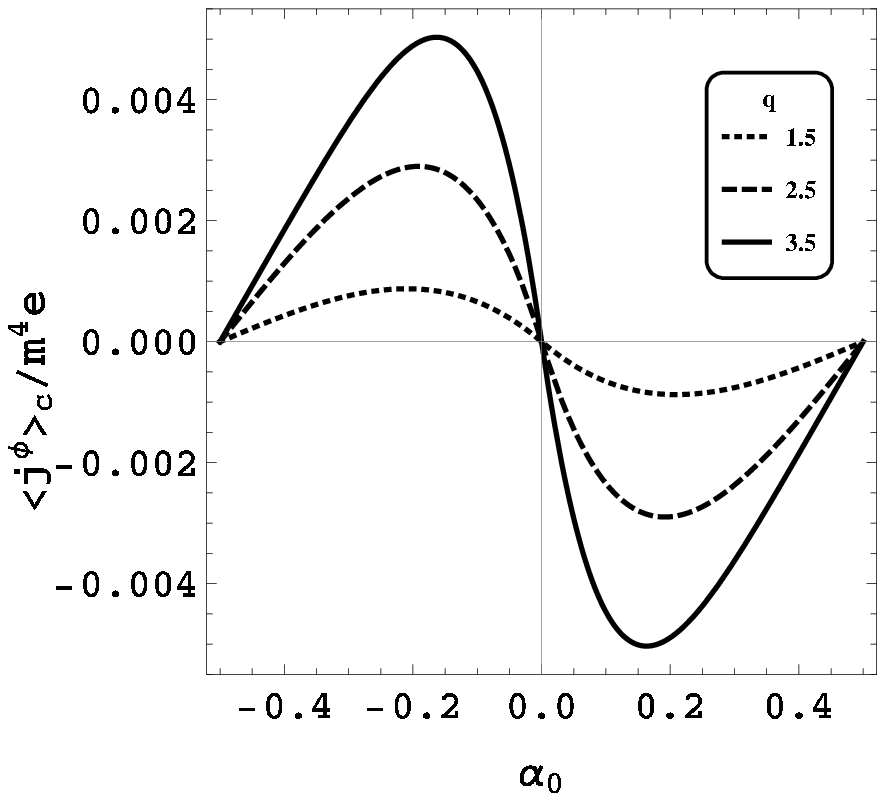}
\caption{The azimuthal current density induced by the compactification for
$D=3$ is plotted, in units of ``$m^4e$'',
in terms of $\alpha_0$ for $mr=0.5$, $mL=1$ and $q=1.5, 2.5$ and 3.5.
The plot on the left is for $\tilde{\beta}=0.1$ while the plot on the
right is for $\tilde{\beta}=0.7$.}
\label{fig2}
\end{center}
\end{figure}

From the above expression we can see that the contribution
due to the compactification on the bosonic current density 
is an even function of the parameter $\tilde{\beta}$
and is an odd function of the magnetic flux along the core of the string, with period equal to $\Phi_0$. In particular,
in the case of an untwisted bosonic field, $\langle j^{\phi }(x)\rangle _{{\rm c}}$
is an even function of the magnetic flux enclosed by the string's axis. Also we can see
that for $\alpha_0=0$ the induced current above vanishes. Moreover, 
$\left\langle j^{\phi}(x) \right\rangle_{{\rm c}}=0$ for $r=0$. This result is in
contrast with the fermionic case where the azimuthal current induced by
the compactification does not vanish at $r=0$, as shown in \cite{SERA}.
In Fig.\ref{fig2} we plot the behavior of the compactified azimuthal
current density as a function of $\alpha_{0}$ for $D=3$, considering
$mr=0.5$, $mL=1$, $\tilde{\beta}=0.1, 0.7$ and different values of the parameter $q$. On the graphs below we can see
that the effect of the cosmic string parameter, $q$, is
to amplify the oscillatory nature of the azimuthal current
with respect to the parameter $\alpha_0$ while the effect of the parameter,
$\tilde{\beta}$, is to change the direction of oscillation as well as diminish
the absolute values of $\left\langle j^{\phi}(x) \right\rangle_{{\rm c}}$.
%
%
%

For large values of the length of the compact dimension, $mL\gg1$, assuming 
that $mr$ is fixed and considering $D=3$, the main contribution comes from the $l=1$ term 
and to leading order we find
\begin{eqnarray}
\left\langle j^{\phi}(x) \right\rangle_{{\rm c}}&\approx &\frac{\sqrt{2} e m^{\frac{3}{2}}
\cos(2\pi\tilde{\beta})e^{-mL}}{\pi^{\frac{9}{2}}L^{\frac{5}{2}}}\left[
\sideset{}{'}\sum_{k=1}^{p}\sin(2k\pi/q)\sin(2k\pi\alpha_{0})\right.\nonumber \\
&&\left. +\frac{q}{\pi}\int_{0}^{\infty}dy\frac{g(q,\alpha_{0},2y) \sinh (2y)}
{\cosh(2qy)-\cos(q\pi)}\right] \ ,
\end{eqnarray}
where we can see that there appear an exponential decay. So in this limit, the contribution to the
total current density is dominated by $\langle j^\phi(x)\rangle_{{\rm cs}}$.

For a massless field and also considering $D=3$, we obtain
\begin{eqnarray}
\left\langle j^{\phi}(x) \right\rangle_{{\rm c}}&=&\frac{4e}{\pi^{2}L^{4}}
\left[\sideset{}{'}\sum_{k=1}^{p}{\sin(2k\pi/q)\sin(2k\pi\alpha_{0})}G_{{\rm c}}({\tilde{\beta}},\rho_k)\right.\nonumber \\
&+&\left.\frac{q}{\pi}\int_{0}^{\infty}dy\frac{g(q,\alpha_{0},2y) \sinh(2y)}{\cosh(2qy)-\cos(q\pi)}
G_{{\rm c}}({\tilde{\beta}},\eta(y))\right] \ ,
\label{Massless}
\end{eqnarray}
where we have defined
\begin{equation}
G_{{\rm c}}(\tilde{\beta},x)=\sum_{l=1}^{\infty }\frac{\cos (2\pi l
\tilde{\beta})}{\left( l^{2}+x^{2}\right) ^{2}} \ .  \label{Cbet}
\end{equation}

The summation above can be developed with the help of \cite{gradshteyn2000table}. So, after some
elementar steps we obtain:
\begin{eqnarray}
G_{{\rm c}}(\tilde{\beta},x) &=&-\frac1{2x^4}+\frac{\pi ^{2}\cosh (2\pi \tilde{\beta}x)}{4x^{2}\sinh
^{2}(\pi x)} \nonumber\\
&+&\pi \frac{\cosh [\pi (1-2\tilde{\beta})x]+2\pi \tilde{\beta}x\sinh [\pi
(1-2\tilde{\beta})x]}{4x^{3}\sinh (\pi x)} \ ,  \label{Cbet1}
\end{eqnarray}
for $0\leqslant \tilde{\beta}\leqslant 1$.

With \eqref{Cbet1} we can also obtain the dominant behavior of $\left\langle j_{\phi}(x) \right\rangle_{{\rm c}}$
in the region $r<<L$. It reads,
\begin{eqnarray}
\left\langle j^{\phi}(x) \right\rangle_{{{\rm c}}}&\approx-&\frac{2e}{45L^{4}}[30
\tilde{\beta}^2(1-\tilde{\beta})^2-1]
\left[\sideset{}{'}\sum_{k=1}^{p}{\sin(2k\pi/q)\sin(2k\pi\alpha_{0})}\right.\nonumber \\
&+&\left.\frac{q}{\pi}\int_{0}^{\infty}dy\frac{g(q,\alpha_{0},2y) \sinh(2y)}{\cosh(2qy)-\cos(q\pi)}
\right] \ .
\end{eqnarray}

Also with \eqref{Cbet1} it is possible to obtain the dominant behavior of 
$\left\langle j^{\phi}(x) \right\rangle_{{\rm c}}$ in the region $r>>L$. Considering
$x>>1$ and $\tilde{\beta}=0$ or $\tilde{\beta}=1$, $G_{{\rm c}}(\tilde{\beta},x)
\approx{\pi}/(4x^3)$. On the other hand taking $x>>1$ and for $0< \tilde{\beta}< 1$,
$G_{{\rm c}}(\tilde{\beta},x)\approx-{1}/(2x^4)$. So we conclude that the
dominant behavior of $\left\langle j^{\phi}(x) \right\rangle_{{\rm c}}$
depends on the value assumed for the parameter $\tilde{\beta}$.
%
\begin{figure}[!htb]
\begin{center}
\includegraphics[width=0.4\textwidth]{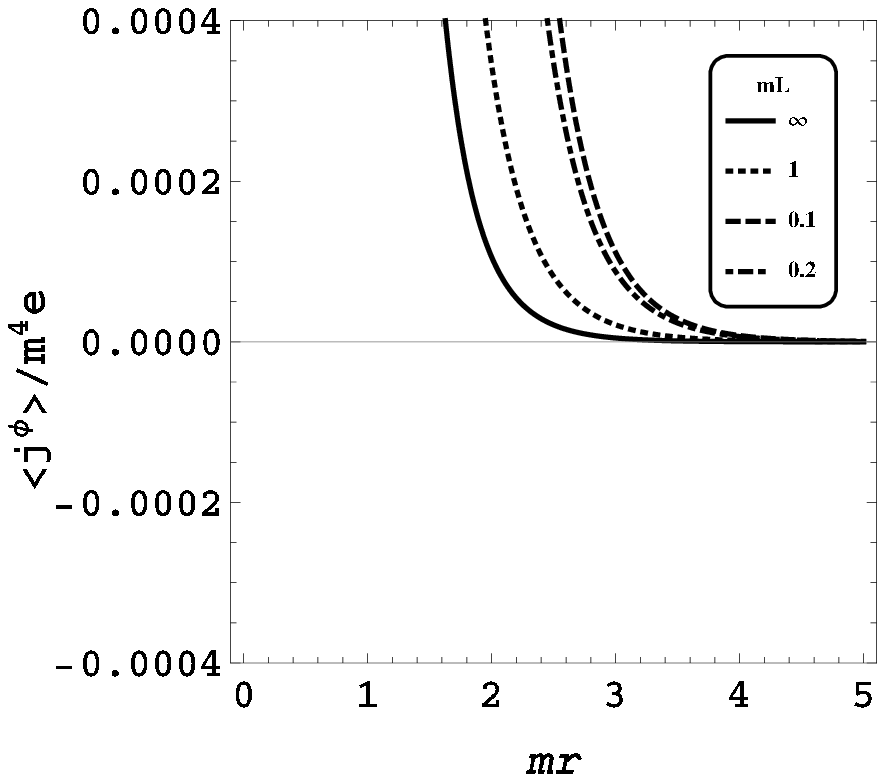}
%
\includegraphics[width=0.4\textwidth]{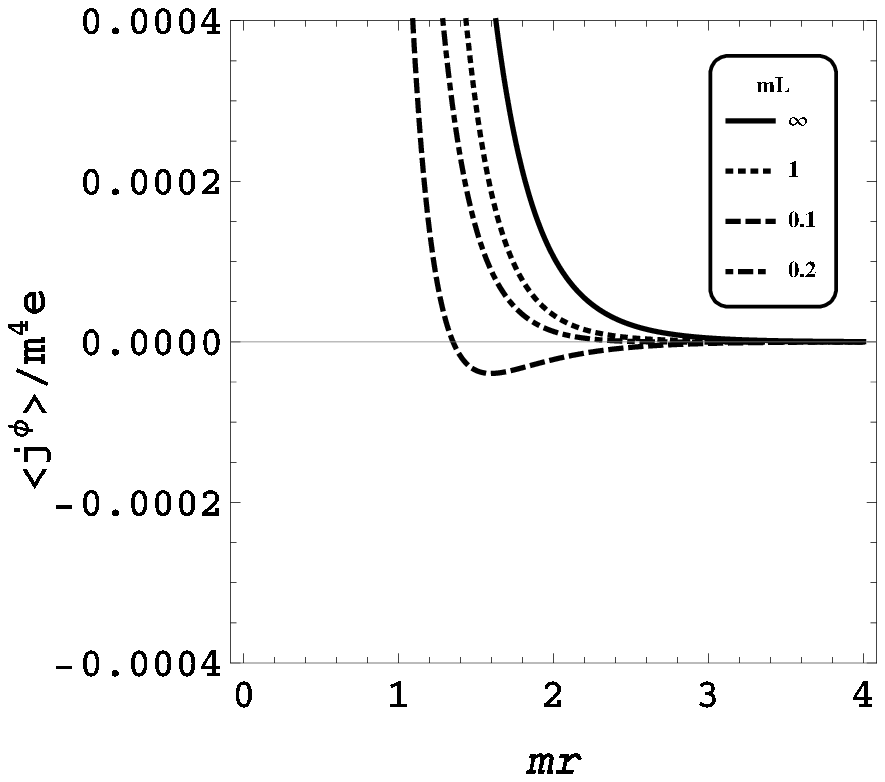}
\caption{The total azimuthal current density for $D=3$ is plotted,
in units of ``$m^4e$'', in terms of $mr$ for the  values $q=2.5$,
$\alpha_0=0.25$ and $mL=0.1,0.2, 1.0$. The curves corresponding
the finite values of $mL$ are compared with the solid curve 
for $mL\rightarrow\infty$, in which case the total azimuthal current
is completely dominated by the term without compactification in
Eq. (\ref{jphi_cs04}). The left plot is for $\tilde{\beta}=0$ while
the right plot is for $\tilde{\beta}=0.5$.}
\label{fig3}
\end{center}
\end{figure}
%

Combining Eqs. \eqref{jphi_cs04} and \eqref{jphi_comp04} we can write the total azimuthal
current as
\begin{eqnarray}
\left\langle j^{\phi}(x) \right\rangle &=&\frac{8em^{D+1}}{(2\pi)^{\frac{(D+1)}{2}}}
\sideset{}{'}\sum_{l=0}^{\infty}
\cos(2\pi l\tilde{\beta})\left\{\sideset{}{'}\sum_{k=1}^{p}\sin(2k\pi/q)
\sin(2k\pi\alpha_{0})F_{\frac{(D+1)}{2}}\left[mL\sqrt{l^{2}+\rho_{k}^{2}}\right]\right.\nonumber \\
&&\left.+ \ \frac{q}{\pi}\int_{0}^{\infty}dy\frac{g(q,\alpha_{0},2y) \sinh{(2y)}
 }{\cosh(2qy)-\cos(q\pi)}F_{\frac{(D+1)}{2}}\left[mL\sqrt{l^{2}+\eta^{2}(y)}\right] \right\} \ ,
\label{jphi_total}
\end{eqnarray}
where the prime on the sum over $l$ means that the term with $l=0$ should be taken with the weight 1/2.
In fig.\ref{fig3} we plot the total azimuthal current density as a function of $mr$ for $D=3$, considering
$q=2.5$, $\alpha_0=0.25$ and different values of $mL$. In the left plot we consider $\tilde{\beta}=0$ while in
the right one $\tilde{\beta}=0.5$. For $A_{z}=0$, one has $\tilde{\beta}=\beta$,
and the cases $\beta=0$ and $\beta=1/2$ are related with the untwisted and twisted bosonic
fields, respectively. One can see that the effect of the parameter $\tilde{\beta}$ is to
alter the curves for the finite values of $mL$ compared with the solid curve for $mL\rightarrow\infty$ .
%
\subsection{Axial current}
\label{axial}
%
Here we shall analyze the bosonic current density along the string. As we shall see,
due to the compactification of the direction along the string axis, there appear
a non-vanishing current. 

The VEV of the axial current is given by
\begin{equation}
\left\langle j_{z}(x) \right\rangle = ie\lim_{x'\rightarrow x}
\left\{(\partial_{z}-\partial_{z'})W(x,x')+2ieA_{z}W(x,x')\right\} \ .
\label{jaxial}
\end{equation}

Using \eqref{eq18} and the fact that $A_z=-\Phi_z/L$, a formal expression for this current can be provided.
It reads,
\begin{equation}
\left\langle j_{z}(x) \right\rangle = -\frac{eq}{L(2\pi)^{D-2}}\sum_{n=-\infty}^\infty
\int d{\vec{k}} \int_0^\infty \lambda \ J^2_{q|n+\alpha|}(\lambda r) 
\sum_{l=-\infty}^\infty\frac {{\tilde{k}}_l}{\sqrt{m^2+\lambda^2+{\tilde{k}}_l+{\vec{k}}}} \ ,
\label{jaxial1}
\end{equation}
where ${\tilde{k}}_l$ is given by \eqref{eq13}.

To evaluate the summation over the quantum number $l$ we shall use the
generalized Abel-Plana summation formula, Eq. \eqref{sumform}. For this case we have
$g(u)=2\pi u/L $ and $f(u)$ is given by \eqref{fg}. Taking these expressions
into consideration, we can see that the first term on the right hand side of Eq. \eqref{sumform}
vanishes due to the fact that $g(u)$ is an odd function. Thus, it
remains only a contribution due to the second term on the right hand side of Eq. \eqref{sumform}.
This contribution
is a consequence of the compactification assumed for
the direction along the cosmic string's axis.

The axial current density induced by the compactification can be written as
\begin{eqnarray}
\left\langle j_{z}(x) \right\rangle_{{\rm c}}&=&-\frac{2ieq}{(2\pi)^{D-1}}
\sum_{n=-\infty}^{\infty}\int d\vec{k}\int_{0}^{\infty}d\lambda \ 
\lambda J_{q|n+\alpha|}^{2}(\lambda r)\nonumber \\
&&\times\int_{\sqrt{\lambda^{2}+\vec{k}^{2}+m^{2}}}^{\infty}\frac{dy \ y}
{\sqrt{y^{2}-\lambda^{2}-\vec{k}^{2}-m^{2}}}\sum_{j=\pm 1}
\frac{j}{e^{Ly-2\pi i j\tilde{\beta}}-1} \ .
\label{ACD}
\end{eqnarray}
Using again the series expansion, $\left(e^{u}-1\right)^{-1}=
\sum_{l=1}^{\infty}e^{-lu}$, for the summation in $j$ present in Eq. \eqref{ACD}, we find
\begin{eqnarray}
\left\langle j_{z}(x) \right\rangle_{{\rm c}}&=&\frac{4eq}{(2\pi)^{D-1}}
\sum_{l=1}^{\infty}\sin(2\pi l\tilde{\beta})\int d{\vec{k}}\int_{0}^{\infty}
d\lambda \ \lambda J_{q|n+\alpha|}^{2}(\lambda r)\nonumber \\
&&\times\int_{\sqrt{\lambda^{2}+{\vec{k}}^{2}+m^{2}}}^{\infty}dy
\frac{y \ e^{-lLy}}{\sqrt{y^{2}-\lambda^{2}-{\vec{k}}^{2}-m^{2}}} \ .
\label{jaxialc}
\end{eqnarray}

The integral over $y$ can be evaluated with help of \cite{gradshteyn2000table}.
The result is given in terms of the modified Bessel function of 
first order, $K_{1}(z)$. On the other hand, using the integral
representation \eqref{Macdonald_repres} for this function, and
the fact that $K_{1}(z)=K_{-1}(z)$ we can show that
\begin{equation}
\int_{\sqrt{\lambda^{2}+\vec{k}^{2}+m^{2}}}^{\infty}dy\frac{y \ e^{-lLy}}
{\sqrt{y^{2}-\lambda^{2}-\vec{k}^{2}-m^{2}}}=\frac{1}{lL}\int_{0}^{\infty}
dt \ e^{-t-\frac{l^{2}L^{2}(\lambda^{2}+\vec{k}^{2}+m^{2})}{4t}} \ .
\end{equation}

Substituting the right hand side of the above identity into
\eqref{jaxialc}, it is possible to develop the integral over the variable 
$\lambda$ and also the integral over the extra momenta.
Moreover, defining a new variable $w=2tr^2/l^2L^{2}$, 
we obtain
\begin{eqnarray}
\left\langle j_{z}(x) \right\rangle_{{\rm c}}&=&\frac{2qeL}{(2\pi)^{\frac{(D+1)}{2}}r^{D+1}}
\sum_{l=1}^{\infty}l\sin(2\pi l\tilde{\beta})\int_{0}^{\infty}dw \ w^{\frac{(D-1)}{2}}
e^{-w\left[1+\frac{l^2L^2}{2r^2}\right]-\frac{m^2r^2}{2w}}\nonumber \\
&&\times\sum_{n=-\infty}^{\infty}I_{q|n+\alpha|}(w) \ .
\label{jaxialc1}
\end{eqnarray}

Substituting the expression involving the summation of the modified Bessel function 
given by \eqref{eq:10} into \eqref{jaxialc1}, we obtain:
\begin{eqnarray}
\left\langle j_{z}(x) \right\rangle_{{\rm c}}&=&\frac{2qeL}{(2\pi)^{\frac{(D+1)}{2}}r^{D+1}}
\sum_{l=1}^{\infty}l\sin(2\pi l\tilde{\beta})\int_{0}^{\infty}dw \ w^{\frac{(D-1)}{2}}
e^{-w\left[1+\frac{l^2L^2}{2r^2}\right]-\frac{m^2r^2}{2w}}\left[\frac{e^{w}}{q}\right.\nonumber \\
&&\left.-\frac{1}{\pi}\int_{0}^{\infty}dy\frac{f(q,\alpha_{0},y)e^{-w\cosh y}}{\cosh(qy)-\cos(q\pi)}+
\frac{2}{q}\sum_{k=1}^{p}\cos(2k\pi\alpha_{0})e^{w\cos(2k\pi/q)}\right] \ ,
\label{jaxial_comp02}
\end{eqnarray}
where we have used $\alpha$ in the form \eqref{alphazero} and $f(q,|\alpha_{0}|,y)$ 
is given by \eqref{eq:09}. 
Integrating over the variable $w$, the above expression can be
written as
\begin{eqnarray}
\left\langle j_{z}(x) \right\rangle_{{\rm c}}=
\left\langle j_{z}(x) \right\rangle_{{\rm c}}^{(0)}+
\left\langle j_{z}(x) \right\rangle_{{\rm c}}^{(q,\alpha_{0})}
\end{eqnarray}

The first term inside the bracket in \eqref{jaxial_comp02}, provides a contribution
that does not depend on $\alpha_0$ and $q$. It is a pure topological term, a consequence of
the compactification only. For this contribution one has
\begin{equation}
\left\langle j^{z}(x) \right\rangle_{{\rm c}}^{(0)}=-\frac{2em^{\frac{(D+1)}{2}}}{(2L)^{\frac{(D-1)}{2}}
\pi^{\frac{(D+1)}{2}}}\sum_{l=1}^{\infty}\frac{\sin(2\pi l\tilde{\beta})}{l^{\frac{(D-1)}{2}}}K_{\frac{(D+1)}{2}}(lmL) \ .
\label{jaxial_compMin}
\end{equation}
Notice that this term is independent on the radial distance, $r$. We can say that the above equation
corresponds to the current density in the $(D+1)$ Minkowski spacetime
with the spatial topology $R^{D-1}\times S$.
The current density in Minkowski spacetime wiht the topology
$R^{p}\times (S^1)^q$ is recently investigated in \cite{deMello:2012xm}
(for the corresponding problem in de Sitter spacetime see \cite{Bellucci:2013vvz}).
Moreover, from \eqref{jaxial_comp02} we can
see that the axial current density vanishes for interger and half-integer values of $\tilde{\beta}$.

For $D=3$, Eq. \eqref{jaxial_compMin} reads
\begin{equation}
\left\langle j^{z}(x) \right\rangle_{{\rm c}}^{(0)}=-\frac{em^2}{\pi^{2}L}
\sum_{l=1}^{\infty}\frac{\sin(2\pi l\tilde{\beta})}{l}K_{2}(lmL) \ .
\end{equation}
This value is exactly the half of the corresponding value for the fermionic case found in \cite{BMSAA}.

\begin{figure}[!htb]
\begin{center}
\includegraphics[width=0.4\textwidth]{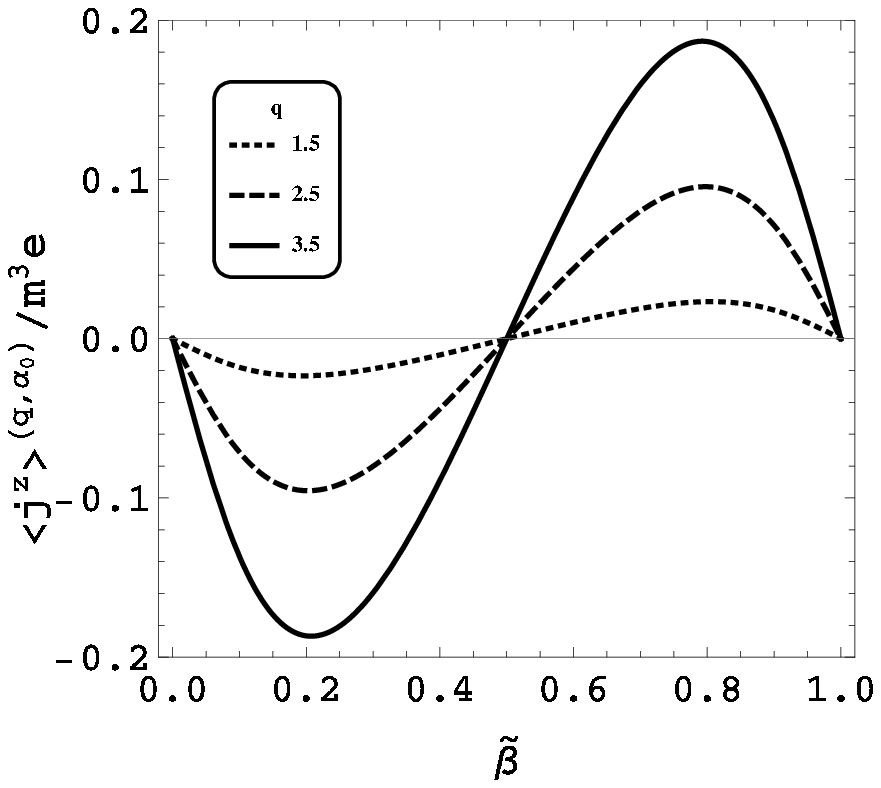}
%
\includegraphics[width=0.398\textwidth]{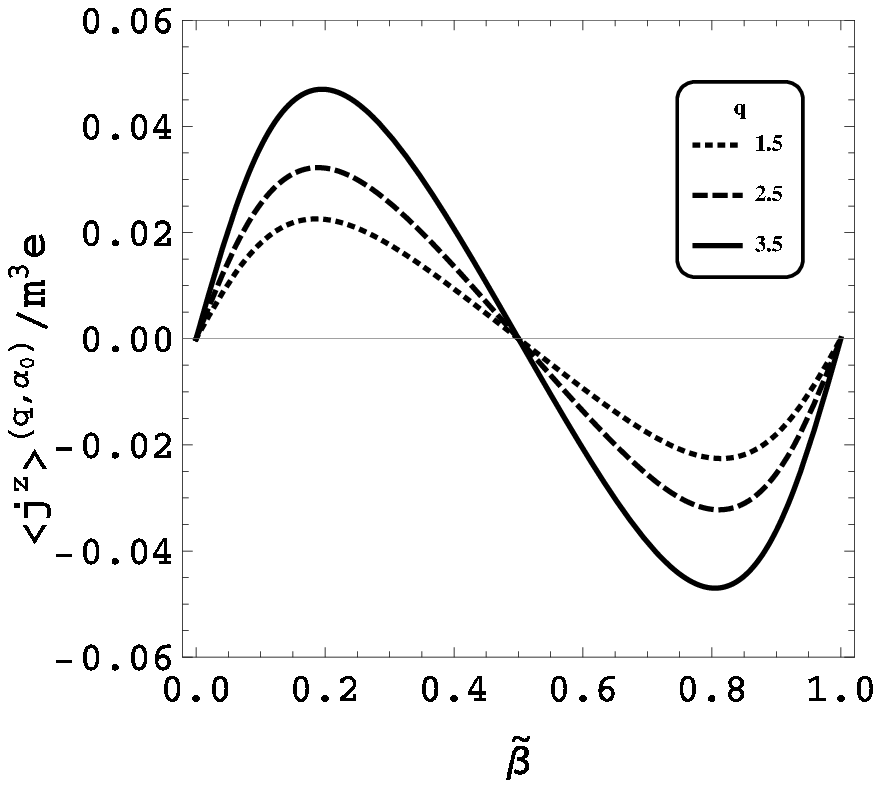}
\caption{The axial current density for $D=3$ in Eq. (\ref{jaxial2_comp}) is plotted,
in units of ``$m^3e$'', in terms of $\tilde{\beta}$ for the values $mr=0.4$,
$mL=1$ and $q=1.5, 2.5, 3.5$. The left plot is for $\alpha_0=0$ while
the right plot is for $\alpha_0=0.25$.}
\label{fig4}
\end{center}
\end{figure}

The second contribution to the axial current coming from the magnetic flux and planar angle deficit is:
\begin{eqnarray}
\left\langle j^{z}(x) \right\rangle_{{\rm c}}^{(q,\alpha_{0})}&=&-\frac{8em^{D+1}L}
{(2\pi )^{\frac{(D+1)}{2}}}\sum_{l=1}^{\infty}l\sin(2\pi l\tilde{\beta})
\left\{\sideset{}{'}\sum_{k=1}^{p}\cos(2k\pi\alpha_{0})F_{\frac{(D+1)}{2}}\left[mL\sqrt{l^{2}+\rho_{k}^{2}}\right]\right.\nonumber \\
&&\left.-\frac{q}{\pi}\int_{0}^{\infty}dy
\frac{f(q,\alpha_{0},2y)}{\cosh(2qy)-\cos(q\pi)}F_{\frac{(D+1)}{2}}\left[mL\sqrt{l^{2}+\eta^{2}(y)}\right]\right\},
\label{jaxial2_comp}
\end{eqnarray}
where $\rho_k$ and $\eta(y)$ are given by \eqref{rhoandeta}. We can see
that Eq. \eqref{jaxial2_comp} is an odd function of the parameter $\tilde{\beta}$ and
is an even function of $\alpha_0$, with period equal to the quantum flux $\Phi_0$. In particular,
in the case of an untwisted bosonic field, Eq. \eqref{jaxial2_comp} is an odd function of the magnetic 
flux enclosed by the string's axis.  Moreover, this contribution vanishes for $q=1$ and $\alpha_0=0$.
In fig.  \ref{fig4} we plot the behavior of $\left\langle j^{z}(x) \right\rangle_{{\rm c}}^{(q,\alpha_{0})}$
as function of ${\tilde{\beta}}$ for $D=3$, considering $mr=0.4$, $mL=1$ and $q=1.5, \ 2.5$ and
$3.5$. The left plot is for $\alpha_0=0$ and the right plot is for $\alpha_0=0.25$. By these plots we can see
that the amplitude of the current increases with $q$ and the effect of $\alpha_0$ is to change the orientation
of the current.
%

For $r=0$ and $D=3$, Eq. \eqref{jaxial2_comp} becomes
\begin{eqnarray}
\left\langle j^{z}(x) \right\rangle_{{\rm c}}^{(q,\alpha_{0})}&=&-\frac{2em^{2}}{\pi^{2}L}
\sum_{l=1}^{p}\frac{K_{2}(lmL)
\sin(2l\pi\tilde{\beta})}{l}\left[\sideset{}{'}\sum_{k=1}^{p}\cos(2k\pi\alpha_{0})\right.\nonumber\\
&-&\left.
\frac{q}{\pi}\int_{0}^{\infty}dy\frac{f(q,\alpha_{0},2y)}{\cosh(2qy)-\cos(q\pi)}\right] \ , 
\label{jaxial2_rzero}
\end{eqnarray}
where we can see that this contribution is finite.

Now, considering $mL\gg1$ and $mr$ fixed, the main contribution to \eqref{jaxial2_comp}
comes from the $l=1$ term. So, for $D=3$ we have
\begin{eqnarray}
\left\langle j^{z}(x) \right\rangle_{{\rm c}}^{(q,\alpha_{0})}& \approx &
-\frac{8em^{\frac{3}{2}}\sin(2\pi\tilde{\beta})e^{-mL}}{(2\pi L)^{\frac{3}{2}}}\left\{
\sideset{}{'}\sum_{k=1}^{p}\cos(2k\pi\alpha_{0})\right.\nonumber\\
&&\left.-\frac q\pi\int_{0}^{\infty}dy\frac{f(q,\alpha_{0},2y)}{\cosh(2qy)-\cos(q\pi)}\right\} \ ,
\end{eqnarray}
where there appears an exponential decay. 

For a massless field and $D=3$, the equation \eqref{jaxial2_comp} reads,
\begin{eqnarray}
\left\langle j^{z}(x) \right\rangle_{{\rm c}}^{(q,\alpha_{0})}&=&-\frac{4e}{\pi^{}L^{3}}
\left[\sideset{}{'}\sum_{k=1}^{p}{\cos(2k\pi\alpha_{0})}V_{{\rm c}}({\tilde{\beta}},\rho_k)\right.\nonumber \\
&&\left. -\frac{q}{\pi}\int_{0}^{\infty}dy\frac{f(q,\alpha_{0},2y)}
{\cosh(2qy)-\cos(q\pi)}V_{{\rm c}}({\tilde{\beta}},\eta(y))\right] \ ,
\end{eqnarray}
where we have defined
\begin{eqnarray}
V_{{\rm c}}({\tilde{\beta}},x)=\sum_{l=1}^\infty\frac{l\sin(2\pi l{\tilde{\beta}})}{(l^2+x^2)^2} \ .
\label{sum-sin}
\end{eqnarray}

Once more with the help of \cite{gradshteyn2000table}, we were able to develop the
summation \eqref{sum-sin}. The result is:
\begin{eqnarray}
V_{{\rm c}}({\tilde{\beta}},x)=-\frac{\pi^2}{4x}\frac{[\sinh(2\pi{\tilde{\beta}}x)-
2{\tilde\beta}\sinh(\pi x)\cosh[\pi x(1-2{\tilde\beta})]]}{\sinh^2(\pi x)} \ ,
\label{sum-sin1}
\end{eqnarray} 
for $0\leqslant \tilde{\beta}\leqslant 1$. Clearly we can see that $V_{{\rm c}}({\tilde{\beta}},x)$ vanishes
for ${\tilde{\beta}}=0, \ 1/2 , \ 1$.

With \eqref{sum-sin1} we can obtain the dominant behavior of $\left\langle j_{z}(x) \right\rangle_{{\rm c}}^{q\neq 1}$
for $r<<L$. It reads,
\begin{eqnarray}
\left\langle j^{z}(x) \right\rangle_{{\rm c}}^{(q,\alpha_{0})}&\approx& -\frac{4e}{L^{3}}\frac{\tilde{\beta}}{3}
(1-\tilde{\beta})(1-2\tilde{\beta})
\left[\sideset{}{'}\sum_{k=1}^{p}{\cos(2k\pi\alpha_{0})}\right.\nonumber \\
&&\left. -\frac{q}{\pi}\int_{0}^{\infty}dy\frac{f(q,\alpha_{0},2y)}
{\cosh(2qy)-\cos(q\pi)}\right] \ .
\end{eqnarray}
In the oposite limit, we can obtain the behavior of $\left\langle j^{z}(x) \right\rangle_{{\rm c}}^{q\neq 1}$
for $r>>L$. Taking the limit $x>>1$ in \eqref{sum-sin1}, we can verify that 
$V_{{\rm c}}({\tilde{\beta}},x)\approx\pi^2{\tilde{\beta}}/(2x)e^{-2\pi{\tilde{\beta}}x}$
for $0<\tilde{\beta}<1/2$, and $V_{{\rm c}}({\tilde{\beta}},x)\approx-\pi^2/(2x)e^{-2\pi(1-{\tilde{\beta}})x}$
for $1/2<\tilde{\beta}<1$. In both cases the axial current presents an exponential decay, however there
appears a different signal.
%
\section{Conclusion}
\label{conc}
%
In this paper we have investigated the bosonic current density
in a higher dimensional compactified cosmic string spacetime
induced by magnetic fluxes, one of them enclosed by the compactified direction
and the other running along the string's core.
The calculations were performed by imposing the quasiperiodicity
condition, with arbitrary phase $\beta$, on the solution of
the Klein-Gordon equation. The general solution was obtained
by considering a constant vector potential in Eq. (\ref{eq05}) and, after imposing the
quasiperiodicity condition and calculating the normalization constant (\ref{eq15}), was presented
in its final form in Eq. (\ref{eq16}). 

The positive frequency Wightman function (\ref{eq18}), which is necessary to calculate the VEV of
the bosonic current density in Eq. (\ref{eq21}), was constructed by the complete
set of normalized wave function  (\ref{eq16}). In this context we were able to show that the
renormalized charge and radial current densities vanish. Moreover, we have seen that the
compactification induces the azimuthal current density to decompose into two parts.
The first one corresponds to the expression in the geometry of a cosmic string without
compactification and is presented in Eq. (\ref{jphi_cs04}). The second contribution is due to the
compactification and is presented in Eq. (\ref{jphi_comp04}).
The former is an odd function of $\alpha_0$, 
with period equal to the quantum flux $\Phi_0$, and is plotted for $D=3$, in units of ``$m^4e$'', 
with respect to $\alpha_0$ as it can be seen in Fig.\ref{fig1}. By this graph we can see
that the intensity of the current increases with the parameter $q$.

Furthermore, we have seen that the azimuthal current density induced by the compactification
is an even function of the parameter $\tilde{\beta}$ and is an odd 
function of the magnetic flux along the core of the string, with period equal to $\Phi_0$. 
We have checked that when $\beta=0$ (untwisted bosonic field),
Eq. (\ref{jphi_comp04}) becomes an even function of the magnetic flux enclosed by the strings axis.
We have also checked that this induced current vanishes for the case $\alpha_0=0$.
In addition, in contrast with the fermionic case investigated in Ref.  \cite{BMSAA},
the azimuthal current density,  $\left\langle j^{\phi}(x) \right\rangle_{{\rm c}}$, vanishes
for $r = 0$. For a massless field and considering $D = 3$, Eq. (\ref{jphi_comp04})
is further simplified and is given by Eq. (\ref{Massless}). In this case, the dominant
behavior of $\left\langle j^{\phi}(x) \right\rangle_{{\rm c}}$ depends on the values assumed
for the parameter $\tilde{\beta}$. In Fig.\ref{fig2}, we plotted Eq. (\ref{jphi_comp04})
for $D=3$, in units of ``$m^4e$'', with respect to $\alpha_0$. Also by this graph
we can see that the intensity of the current increases with $q$, and the effect
of the parameter $\tilde{\beta}$ plays an important rule on the sign of direction.

For the total azimuthal current density, that is, the sum of Eqs. (\ref{jphi_cs04})
and (\ref{jphi_comp04}), we have seen that it is dominated by
$\left\langle j^{\phi}(x) \right\rangle_{{\rm cs}}$ for large values
of the length of the compact dimension, $mL\gg 1$, assuming
$D=3$ and $mr$ fixed. A plot of the total azimuthal
current density, in units of ``$m^4e$'', with respect to $mr$ is presented in Fig.\ref{fig3}
for $D = 3$ and for two different values of $\tilde{\beta}$. From this graph we can see that the relavance
of the compactified part of the current depends on the product $mL$, decreasing when $mL$
becomes larger. Moreover, the relative intensity of the total current, compared
with the $\left\langle j^{\phi}(x) \right\rangle_{{\rm cs}}$,
depends on $\tilde{\beta}$.

We have also shown that the VEV of the axial current density in
Eq. (\ref{jaxial_comp02}) has a purely topological origin and
vanishes when $\tilde{\beta}=0,1/2$ and $1$. This VEV can be
expressed as the sum of two terms.
One of them is given by Eq. (\ref{jaxial_compMin}) and independ of the
radial distance $r$, the cosmic string parameter $q$ and $\alpha_0$.
This contribution corresponds to the current density
in $(D+1)$ Minkowski spacetime with the spatial
topology $R^{D-1}\times S$. The other contribution is given by
Eq. (\ref{jaxial2_comp}) and is due to the magnetic fluxes
and the planar angle deficit. We verified that this contribution is an 
odd function of the parameter $\tilde{\beta}$ and is an even 
function of $\alpha_0$, with period equal to the quantum 
flux $\Phi_0$. For the particular case when $\beta=0$, 
Eq. (\ref{jaxial2_comp}) becomes an odd function of the magnetic flux enclosed by the strings axis.
A plot of the azimuthal current as function ${\tilde{\beta}}$ is presented in Fig.\ref{fig4}
for two different values of $\alpha_0$ and considering $D=3$. By this graph we can see 
that the amplitude of the current increases with the parameter $q$ and the effect of $\alpha_0$ is 
to change the orientation of the current.

We would like to emphasize that the currents analyzed in this paper
refer to the vacuum ones, induced by the presence of the magnetic fluxes
and the compactification. As we could see, the planar angle deficit associated with the
cosmic string spacetime amplifies the oscillatory nature of the azimuthal current density, 
and the compactification introduces additional contribution to it, 
creating a new axial current density. 
Additionally to the vacuum induced current, Witten has shown that under specific
conditions, cosmic strings behave as  superconducting wires  \cite{Witten1985557}, being 
the particle excitation  the superconducting charge carriers, that may be 
either bosons or fermions (see also \cite{Carter:1999hx,Ringeval:2001xd}).  
Moreover, in the context of braneworlds, the stability of superconducting cosmic strings 
was discussed in \cite{Peter:2003zg}, and  the  equations of states relating the tension and energy 
density per unit length of such topological defects was investigated in \cite{PhysRevD.45.1091}.

%
\section*{Acknowledgments}
The authors E.A.F.B and H.F.S.M thank the Brazilian agency CAPES for financial support.
E.R.B.M. thanks Conselho Nacional de Desenvolvimento Cient\'{\i}fico e Tecnol\'ogico (CNPq)
for partial financial support. Also we want to thank Aram A. Saharian for a critical reading of the
paper.
%
\appendix

\section{Summation formulas}
\label{summ}
Here, we shall develop the summation involving the modified Bessel functions in Eqs. \eqref{sum1} and \eqref{jphi_cs03}. 

\subsection{Summation formula involving the modified Bessel function $I_{\beta_n}(w)$}
\label{summ1}
We start first with the expression \eqref{sum1}. Let us consider the sum
\begin{eqnarray}
\mathcal{I}(w,\alpha,q)=\sum_{n=-\infty}^{\infty}I_{q|n+\alpha|}(w)=
I_{q|\alpha_0|}(w)+\sum_{n=1}^{\infty}[I_{q(n+\alpha_0)}(w)+I_{q(n-\alpha_0)}(w)].
\label{eq:01}
\end{eqnarray}
A very useful integral representation for $I_{\beta_n}(w)$ has previously 
been considered in \cite{PhysRevD.82.085033} and will also be used here. This representation is given by
\begin{eqnarray}
I_{\beta_n}(w)=\frac{\sin(\pi\beta_n)}{\pi\beta_n}e^{-w}+\frac{w}{\pi}
\int_{0}^{\pi}dy\sin y\frac{\sin(y\beta_n)}{\beta_n}e^{w\cos y}-
\frac{\sin(\pi\beta_n)}{\pi}\int_{0}^{\infty}
dye^{-w\cosh y-\beta_ny},
\label{eq:02}
\end{eqnarray}
where in our case $\beta_n=q|n+\alpha_0|$. Upon substituting Eq. (\ref{eq:02}) into 
Eq. (\ref{eq:01}), one can work out each term separately.  Thereby, using Eq. (06) 
(subsection 5.4.3) from \cite{prudnikov}, the summation 
in $n$ of the first and second terms on the right hand side of (\ref{eq:02}) is found to be
\begin{eqnarray}
\sum_{n=-\infty}^{\infty}\frac{\sin(\beta_n\theta)}{\beta_n}=
\frac{\pi}{q\sin(\pi\alpha_0)}\sin[(2k+1)\pi\alpha_0].
\label{eq:05}
\end{eqnarray}
which is only valid for $2k\pi/q<\theta<(2k+2)\pi/q$. 

Now, lets us consider the summation in $n$ of the last term on the right hand side of (\ref{eq:02}). This summation can be rewritten as
\begin{eqnarray}
\sum_{n=-\infty}^{\infty}\sin(\pi\beta_n)e^{-\beta_ny}=\sin(\pi q\alpha_0)
e^{-q\alpha_0y}&+&\sum_{n=1}^{\infty}\left[\sin[(n+\alpha_0)q\pi]e^{-(n+\alpha_0)qy}\right.\nonumber \\
&&+\left.\sin[(n-\alpha_0)q\pi]e^{-(n-\alpha_0)qy}\right].
\label{eq:06}
\end{eqnarray}
In addition, one can further consider the sum on the right hand side of (\ref{eq:06}) in the following form:
\begin{eqnarray}
e^{\mp\alpha_0qy}\sum_{n=1}^{\infty}\sin[(n\pm\alpha_0)q\pi]e^{-nqy}&=
&e^{\mp\alpha_0qy}\left[\cos(\alpha_0\pi q)\sum_{n=1}^{\infty}
\sin(n\pi q)e^{-nqy}\right.\nonumber \\
&&\left.\pm\sin(\alpha_0\pi q)\sum_{n=1}^{\infty}\cos(n\pi q)e^{-nqy}\right].
\label{eq:07}
\end{eqnarray}
Thus, we can use Eqs. (01) and (02) (subsection 5.4.12) from \cite{prudnikov} 
to perform the sums on the right hand side of Eq. (\ref{eq:07}). After doing 
so, and substituting the result in Eq. (\ref{eq:06}), we obtain
\begin{eqnarray}
\sum_{n=-\infty}^{\infty}\sin(\pi\beta_n)e^{-\beta_ny}=\frac{f(q,\alpha_0,y)}
{\cosh(qy)-\cos(\pi q)},
\label{eq:08}
\end{eqnarray}
where
\begin{eqnarray}
f(q,\alpha_0,y)=\sin[(1-|\alpha_0|)\pi q]\cosh(|\alpha_0| qy)+
\sin(|\alpha_0|\pi q)\cosh[(1-|\alpha_0|)qy].
\label{eq:09}
\end{eqnarray}
As it can be seen in Eq. (\ref{eq:09}), we are considering only absolute values of the parameter $\alpha_0$.
The absolute value $|\alpha_0|$ is necessary in order to make Eq. (\ref{eq:09}) an even function
of $\alpha_0$ and, therefore, compatible with Eq. (\ref{eq:01}) which is also an even function
of the same parameter.

Finally, combining Eqs. (\ref{eq:01}), (\ref{eq:02}), (\ref{eq:05}) and (\ref{eq:08}), we are able to show that
\begin{eqnarray}
\mathcal{I}(w,\alpha_{0}, q)&=&\frac{e^w}{q}-\frac{1}{\pi}
\int_{0}^{\infty}dy\frac{e^{-w\cosh y}f(q,\alpha_0,y)}{\cosh(qy)-\cos(\pi q)}\nonumber\\
&+&\frac{2}{q}\sideset{}{'}\sum_{k=1}^{[q/2]}\cos(2k\pi\alpha_0)e^{w\cos(2k\pi/q)} \ ,
\label{eq:10}
\end{eqnarray}
where $[q/2]$ represents the integer part of $q/2$, and the prime on the sign of
the summation means that in the case $q=2p$ the term $k=q/2$ should be
taken with the coefficient $1/2$. Note that by taking the summation in $k$ from $-p$ to 
$+p$ in  (\ref{eq:05}) it provides the last term on the right hand side of (\ref{eq:10}) 
after combining Eqs. (\ref{eq:01}), (\ref{eq:02}), (\ref{eq:05}) and (\ref{eq:08}).
We can see now that Eq. (\ref{eq:10}) is in perfect agreement with Eq. (\ref{eq:01}),
i.e, both are even functions of $\alpha_0$. This is only possible by
considering the absolute values of $\alpha_0$ in Eq. (\ref{eq:09}). 

For integer values of $q$ and $\alpha_0=0$, we have:
\begin{eqnarray}
\mathcal{I}(w, q)=\frac{e^w}{q}+\frac{1}{q}
\sum_{k=1}^{q-1}e^{w\cos(2k\pi/q)}.
\label{eq:12}
\end{eqnarray}
The special case in Eq. (\ref{eq:12}) has been considered in a number of other contexts (see, e.g., \cite{SBM, SBM2, SBM3, Spinelly200477, SBM4}).
%

\subsection{A second summation formula involving the modified Bessel function $I_{\beta_n}(z)$}
\label{summ2}

Let us now turn to the proof of the summation in $n$ presented in Eq. \eqref{jphi_cs03}, which is considered here
in the following form:
\begin{eqnarray}
S=\sum_{n=-\infty}^\infty(n+\alpha_0)I_{q|n+\alpha_0|}(w)=\alpha_0I_{q|\alpha_0|}(w)+
\sum_{n\geq 1}\left[(n+\alpha_0)I_{q(n+\alpha_0)}(w)-(n-\alpha_0)I_{q(n-\alpha_0)}(w)\right],\nonumber\\
\label{form_grad}
\end{eqnarray}
which is an odd function of the parameter $\alpha_0$. Thus, taking firstly
only positive values of $\alpha_0$ and using the 
relations \cite{gradshteyn2000table}:
\begin{eqnarray}
(n+\alpha_0)I_{q(n+\alpha_0)}(w)&=&-\frac wq\frac d{dw}I_{q(n+\alpha_0)}(w)
+\frac wq I_{q(n+\alpha_0)-1}(w) \ , \nonumber\\
(n-\alpha_0)I_{q(n-\alpha_0)}(w)&=&\frac wq\frac d{dw}I_{q(n-\alpha_0)}(w)
-\frac wq I_{q(n-\alpha_0)+1}(w) \ ,
\label{S}
\end{eqnarray}
in Eq. (\ref{form_grad}), we get
\begin{eqnarray}
S&=&-\frac wq\frac d{dw}\left[I_{q|\alpha_0|}(w)+\sum_{n\geq 1}I_{q(n+\alpha_0)}(w)
+\sum_{n\geq 1}I_{q(n-\alpha_0)}(w)\right]\nonumber\\
&&+\frac wq \left[I_{q|\alpha_0|-1}(w)
+\sum_{n\geq 1}I_{q(n+\alpha_0)-1}(w)
+\sum_{n\geq 1}I_{q(n-\alpha_0)+1}(w)\right] \ .
\label{S1}
\end{eqnarray}
Upon defining, $\tilde{\alpha}_0=\alpha_0-1/q$, we are able to rewrite (\ref{S1}) as
\begin{eqnarray}
S&=&-\frac wq\frac d{dw}\sum_{n=-\infty}^\infty I_{q|n+{\alpha}_0|}(w)
+\frac wq \sum_{n=-\infty}^\infty I_{q|n+\tilde{\alpha}_0|}(w).
\label{soma1}
\end{eqnarray}
On the other hand, we can also exchange the factors $(n\pm\alpha)$ between the relations presented in Eq. \eqref{S}. This provides
\begin{eqnarray}
S=\frac wq\frac d{dw}\sum_{n=-\infty}^\infty I_{q|n+{\alpha}_0|}(w)-
\frac wq \sum_{n=-\infty}^\infty I_{q|n+\tilde{\alpha}_0|}(w) \ ,
\label{S2}
\end{eqnarray} 
where now, $\tilde{\alpha}_0=\alpha_0+1/q$. Substituting (\ref{eq:10}) 
into (\ref{S2}), one can separate the latter in two terms, i.e
\begin{eqnarray}
S=S_{\alpha_0}+S_{\tilde{\alpha}_0}
\label{soma}
\end{eqnarray}
The first term is given by
\begin{eqnarray}
S_{\alpha_0}&=&-\frac{2w}{q^2}\sideset{}{'}\sum_{k=0}^{[q/2]}\cos(2k\pi/q)
\cos(2k\pi\alpha_0)e^{w\cos(2k\pi/q)}\nonumber \\
&&-\frac{w}{q\pi}\int_{0}^{\infty}dy\cosh y\frac{e^{-w\cosh y}
f(q,\alpha_0,y)}{\cosh(qy)-\cos(\pi q)},
\label{s1}
\end{eqnarray}
and the second term, $S_{\tilde{\alpha}_0}$, is given in terms of $S_{\alpha_0}$ as shown below:
\begin{eqnarray}
S_{\tilde{\alpha}_0}&=&-S_{\alpha_0}+\frac{2w}{q^2}
\sideset{}{'}\sum_{k=1}^{[q/2]}\sin(2k\pi/q)\sin(2k\pi\alpha_0)e^{w\cos(2k\pi/q)}\nonumber\\
&&+\frac{w}{q\pi}\int_{0}^{\infty}dy\sinh y
\frac{e^{-w\cosh y}g(q,\alpha_0,y)}{\cosh(qy)-\cos(\pi q)},
\label{s2}
\end{eqnarray}
where the function, $g(q,\alpha_0,y)$, is defined as
\begin{eqnarray}
g(q,\alpha_0,y)=\sin(q\pi\alpha_0)
\sinh[(1-|\alpha_0|)qy]-\sinh(yq\alpha_0)\sin[(1-|\alpha_0|)\pi q].
\label{s3}
\end{eqnarray}
As we took only positive values of $\alpha_0$, Eq. (\ref{s3}) does not need to depend on $|\alpha_0|$.
However, if we took negative values of $\alpha_0$ in Eq. (\ref{form_grad}) we would get a similar
expression as in Eq. (\ref{s3}) but with the opposite sign. This suggests that in order to
satisfy both possibilities one has to include absolute values of $\alpha_0$ as showed in Eq. (\ref{s3}).

Therefore, substituting (\ref{s2}) into (\ref{soma}) we finally arrive at
\begin{eqnarray}
S=\frac{2w}{q^2}\sideset{}{'}\sum_{k=1}^{[q/2]}\sin(2k\pi/q)\sin(2k\pi\alpha_0)
e^{w\cos(2k\pi/q)}+\frac{w}{q\pi}\int_{0}^{\infty}dy\sinh y
\frac{e^{-w\cosh y}g(q,\alpha_0,y)}{\cosh(qy)-\cos(\pi q)}.\nonumber\\
\label{s4}
\end{eqnarray}
Note that both expression in Eqs. (\ref{form_grad}) and Eq. (\ref{s4}) are odd function of $\alpha_0$,
as it should be. This is only possible if we take absolute values of $\alpha_0$ as showed in Eq. (\ref{s3}).
Note also that by using Eq. (\ref{soma1}), instead of Eq. (\ref{S2}), 
we would obtain the same expression (\ref{s4}).

\section{Analysis of the induced current densities along the extra dimensions}
\label{extra-dimension}

The VEV of the current densities along the extra dimensions, $\langle j^i(x)\rangle$
for $i= 4, \ \cdots \ D$, are given by
\begin{eqnarray}
\langle j_i(x)\rangle= ie\lim_{x' \rightarrow x}
(\partial_{x^i}-\partial_{x'^i})W(x,x') \ ,
\label{jextra}
\end{eqnarray}
which can be written as
\begin{eqnarray}
\langle j_i(x)\rangle&=&-\frac{eq}{L(2\pi)^{D-2}}\lim_{x^{'} \rightarrow x}
\sum_\sigma e^{iqn\Delta\phi} 
 e^{i{\vec{k}\cdot{\Delta\vec{r}}_{\parallel}}} k^i\ 
\lambda J_{q|n+\alpha|}(\lambda r) J_{q|n+\alpha|}(\lambda r')\nonumber\\
&\times&\frac{e^{-i\omega_l\Delta t+ik_l\Delta z}}{\omega_l} \ .
\label{jextra1}
\end{eqnarray}

The integral over the variable $k^i$ can be evaluated by using the
Eq. \eqref{eq12} and the identity 
\begin{equation}
\frac{1}{\sqrt{m^2+\lambda^2+{\tilde{k}}^2_l+{\vec{k}}^2}}=\frac{2}
{\sqrt{\pi}}\int_{0}^{\infty}ds \ e^{-(m^2+\lambda^2+{\tilde{k}}^2_l+{\vec{k}}^2)s^{2}} \ .
\label{ident}
\end{equation}

Then, the integral in $k^i$ reads
\begin{equation}
\int_{-\infty}^\infty \ dk^i \ k^i e^{-s^2(k^i)^2}e^{ik^i\Delta x^i_{\parallel}}
=\frac{i{\sqrt\pi}\Delta x^i_{\parallel}}{2s^3}e^{-\frac{(\Delta x^i_{\parallel})^2}{4s^2}} \ .
\label{eq-aux}
\end{equation}

This expression goes to zero by taking the limit, $\Delta x^i_{\parallel}\rightarrow 0$.
Moreover, at the coincidence limit, $r' \rightarrow r$, the integral over 
$\lambda$ provides a result equivalent to \eqref{intJ}. In addition the integrals over 
the other components of momentum along the extra dimensions, $k^r$ with $r\neq i$,
provide finite results similar to
the one given by \eqref{intK}. Using again the results
\eqref{eq:10} and \eqref{eq:09} for the sum over the modified Bessel function and
identifying the first term with the contribution of the Minkowski space
 in the absence of magnetic flux, we can renormalize these current densities in a manifest form by
discarding this term. By doing this, we can see that the terms inside the
summation and integral contain factors $e^{-r^2\sin^2(\pi k/q)/s^2}$ and $e^{-r^2\cosh^2(y/2)/s^2}$,
respectively. Consequently, 
the integrals over $s$ of the remaining terms are finite. So, our final conclusion is that, because 
\eqref{eq-aux} goes to zero at the coincidence limit and the renormalized values for the other integrals
are finite in that limit, there will be no induced vacuum current densities along the 
extra dimensions.

%
\begingroup\raggedright\endgroup

\end{document}